\documentstyle[prl,aps,epsfig,amssymb]{revtex}
\def\figwidth{13cm}
\begin{document}
\newcommand{\beq}{\begin{equation}}\newcommand{\eeq}{\end{equation}}
\newcommand{\barr}{\begin{eqnarray}}\newcommand{\earr}{\end{eqnarray}}

\newcommand{\andy}[1]{ }

\def\txt{\textstyle}

\def\ask{\marginpar{?? ask:  \hfill}}
\def\fin{\marginpar{fill in ... \hfill}}
\def\note{\marginpar{note \hfill}}
\def\check{\marginpar{check \hfill}}
\def\discuss{\marginpar{discuss \hfill}}
\def\hh{\widehat}
\def\wtilde{\widetilde}
\newcommand{\bm}[1]{\mbox{\boldmath $#1$}}
\newcommand{\bmsub}[1]{\mbox{\boldmath\scriptsize $#1$}}
\newcommand{\bmh}[1]{\mbox{\boldmath $\hat{#1}$}}

 \def\ch{\mbox{ch}}
 \def\sh{\mbox{sh}}
 \newcommand{\ket}[1]{| #1 \rangle}
 \newcommand{\bra}[1]{\langle #1 |}

\draft

\title{ From the quantum Zeno to the inverse quantum Zeno effect }
\author{ P. Facchi,$^{1}$ H. Nakazato$^{2}$ and S. Pascazio$^{1}$ }

\address{$^{1}$Dipartimento di Fisica, Universit\`a di Bari
     and Istituto Nazionale di Fisica Nucleare, Sezione di Bari,
 I-70126 Bari, Italy \\
$^{2}$Department of Physics, Waseda University,  Tokyo 169-8555,
Japan }

\date{\today}

\maketitle

\begin{abstract}
The temporal evolution of an unstable quantum mechanical system
undergoing repeated measurements is investigated. In general, by
changing the time interval between successive measurements, the decay
can be accelerated (inverse quantum Zeno effect) or slowed down
(quantum Zeno effect), depending on the features of the interaction
Hamiltonian. A geometric criterion is proposed for a transition to
occur between these two regimes.
\end{abstract}


The temporal evolution of the survival probability of a quantum
mechanical unstable system is characterized by a short-time quadratic
behavior, an intermediate approximately exponential decay and a
long-time power tail \cite{temprevi}. The short-time region has
attracted the attention of physicists since quite some time ago,
because it leads, under particular conditions, to the quantum Zeno
effect (QZE)
\cite{Beskow}, by which frequent observations slow down the
evolution. However, it has recently been pointed out that by
exploiting the short-time features of the quantal evolution one can
also accelerate the decay \cite{FPOlomouc,KKNature,perina,LewRza}. We
will call this phenomenon inverse quantum Zeno effect (IZE).

In this Letter we shall analyze how the Zeno--inverse Zeno
transition takes place when the frequency of observations is
changed. For an oscillating quantum mechanical system, whose
Poincar\'e time is finite, it is not difficult to obtain a QZE.
On the other hand, when the system is unstable, the situation is
much more interesting and involved: in general, one can obtain
both a QZE or an IZE depending on the features of the interaction
Hamiltonian.

Let us summarize the main features of the QZE. Prepare, at $t=0$, a
quantum system in some (normalizable) initial state. A QZE typically
arises if one performs a series of ``measurements," at time intervals
$\tau$, in order to ascertain whether the system is still in its
initial state. If $P(t)$ denotes the undisturbed survival probability
in the initial state, after the $N$th measurements the survival
probability reads
 \andy{effgamma}
\beq
P^{(N)}(t)=P(\tau)^N\equiv \exp[-\gamma(\tau)t],
 \label{eq:effgamma}
\eeq
where $t=N\tau$ is the total duration of the experiment and we have
introduced an {\em effective} decay rate $\gamma(\tau)$, which is
defined through the last equality. Notice that the far r.h.s.\
represents an exponential ``interpolation" of $P^{(N)}(t)$ and that
$\gamma$ is in general $\tau$ dependent: for example, if the short
time behavior is $P(t)
\simeq
\exp (-t^2/\tau_{\rm Z}^2)$, where $\tau_{\rm Z}$ is the so-called
Zeno time, one easily checks that $\gamma(\tau)\simeq \tau/\tau_{\rm
Z}^2$. Moreover, one expects to recover the ``natural" lifetime
$\gamma_0^{-1}$, in agreement with the Fermi ``golden" rule, for
sufficiently long time intervals $\tau$. Equation (\ref{eq:effgamma})
is valid $\forall t=N\tau$ and therefore, in particular, for $t=\tau$
(namely, when a single measurement is performed: $N=1$). Hence
\andy{effgammaone}
\beq
\exp[-\gamma(\tau)\tau]= P(\tau) = |x(\tau)|^2 ,
\label{eq:effgammaone}
\eeq
so that $\gamma(\tau)$ is the decay rate of an exponential curve that
intersects the undisturbed survival probability exactly at time
$\tau$ \cite{temprevi,contvspulsed}. In Eq.\ (\ref{eq:effgammaone}),
$x(t)$ is the survival amplitude in the initial state. From Eq.\
(\ref{eq:effgammaone}) one gets the handy formula
\andy{gammadef}
\beq
\label{eq:gammadef} \gamma(\tau)=-\frac{1}{\tau}\ln P(\tau)
=-\frac{2}{\tau}\ln|x(\tau)| =-\frac{2}{\tau}{\rm Re} [\ln x(\tau)],
\eeq
expressing the effective lifetime in terms of the free survival
probability or amplitude.

\begin{figure}
\centerline{\epsfig{file=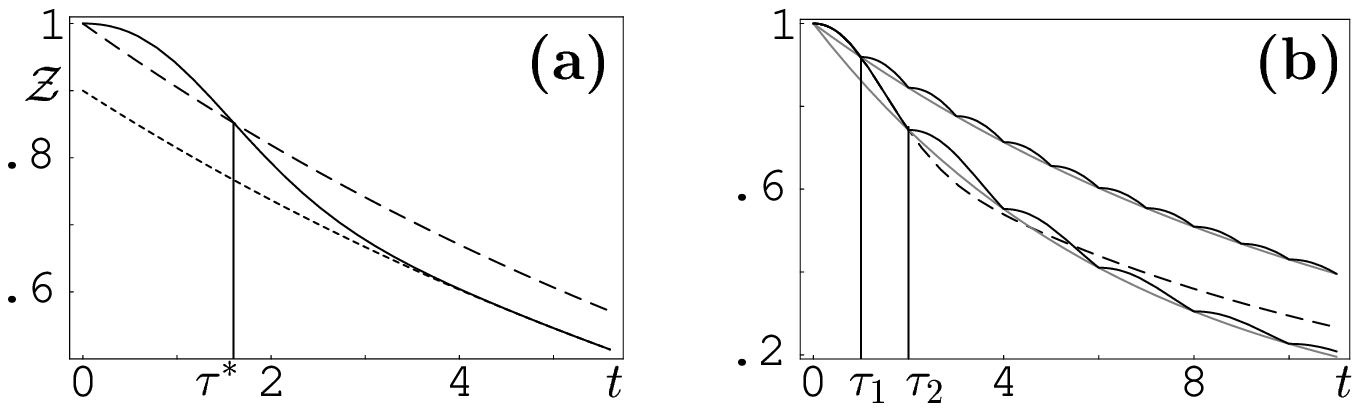,width=\figwidth}}
\caption{(a) Determination of $\tau^*$. The full line is the
survival probability, the dashed line the exponential
$e^{-\gamma_0 t}$ and the dotted line the asymptotic exponential
${\cal Z} e^{-\gamma_0 t}$ [see Eq.\ (\ref{eq:psurv}) in the
following]. (b) Quantum Zeno vs inverse Zeno effect. The dashed
line represents a typical behavior of the survival probability
$P(t)$ when no measurement is performed: the short-time Zeno
region is followed by an approximately exponential decay with a
natural decay rate $\gamma_0$. When measurements are performed at
time intervals $\tau$, we get the effective decay rate
$\gamma(\tau)$. The full lines represent the survival
probabilities and the dotted lines their exponential
interpolations, according to (\ref{eq:effgamma}). For
$\tau_1<\tau^*<\tau_2$ the effective decay rate $\gamma(\tau_1)$
[$\gamma(\tau_2)$] is smaller (QZE) [larger (IZE)] than the
``natural" decay rate $\gamma_0$. When $\tau=\tau^*$ one recovers
the natural lifetime, according to (\ref{eq:tstardef}).}
\label{fig:gtau}
\end{figure}

We now ask whether it is possible to find a time $\tau^*$ such that
\andy{tstardef}
\beq
\gamma(\tau^*)=\gamma_0.
\label{eq:tstardef}
\eeq
If such a time exists, then by performing measurements at time
intervals $\tau^*$ the system decays according to its ``natural"
lifetime, as if no measurements were performed. Figure~\ref{fig:gtau}
illustrates an example in which such a time exists: if the curves
$e^{-\gamma_0t}$ and $P(t)$ intersect, their intersection is at
$\tau^*$. (Notice that there can be more than one intersection, i.e.\
Eq.\ (\ref{eq:tstardef}) can have more solutions, e.g.\ if $P(t)$
oscillates around $e^{-\gamma_0t}$ \cite{hydrovanH}. In such a case,
$\tau^*$ is defined as the smallest solution.) It is apparent that if
$\tau<\tau^*$ one obtains a QZE. {\it Vice versa}, if $\tau>\tau^*$,
one obtains an {\it inverse} Zeno effect. In this sense, $\tau^*$ can
be viewed as a {\it transition time} from a quantum Zeno to an
inverse Zeno regime. Paraphrasing Misra and Sudarshan
\cite{Beskow}, we can say that $\tau^*$ determines the transition
from Zeno (who argued that a sped arrow, if observed, does not move)
to Heraclitus (who replied that everything flows). We shall see that
in general it is not always possible to determine $\tau^*$: Eq.\
(\ref{eq:tstardef}) may have no finite solutions. This depends on
several features of the evolution law and will be discussed in the
following.

We shall work in a quantum field theoretical framework. Consider the
Hamiltonian ($\hbar =1$)
\andy{contham}
\beq
H=H_0+H_{\rm int} = \omega_a\ket{a}\bra{a} +\int d\omega\;\omega
\ket{\omega} \bra{\omega} +\int d\omega\;g(\omega) ( \ket{a}
\bra{\omega}+ \ket{\omega} \bra{a}),
\label{eq:contham}
\eeq
where $\langle a|a\rangle=1, \langle a|\omega \rangle=0$ and $
\langle \omega|\omega'\rangle=\delta(\omega-\omega')$. It describes
the interaction of a normalizable (discrete) state $|a \rangle$ (the
initial state) with a continuum of states $|\omega \rangle$ into
which it can decay; $g(\omega)$ is the form factor of the
interaction. The survival amplitude and probability of finding the
system still in the initial state $|a\rangle$ at $t>0$ read
\andy{xPt}
\beq
x(t) \equiv \langle a|\psi(t)\rangle, \qquad P(t)=|\langle
a|\psi(t)\rangle|^2=|x(t)|^2,
\label{xPt}
\eeq
respectively, where $|\psi(t) \rangle$ is the state at time $t$,
whose evolution is naturally restricted to the Tamm-Duncoff sector
spanned by $\{\ket{a}, \ket{\omega}\}$. The survival amplitude is
conveniently written as the inverse Fourier-Laplace transform of the
propagator $x(E)=i\bra{a}(E-H)^{-1}\ket{a}$,
\andy{antitran}
\beq
\label{eq:antitran}
x(t)=\int_{\rm B}\frac{dE}{2\pi}\;e^{-iEt}x(E), \qquad
x(E)=\frac{i}{E-\omega_a-\Sigma(E)},
\eeq
where the Bromwich path B is a horizontal line $\mbox{Im}
E=$const$>0$ in the half plane of analyticity of the transform
(upper half plane) and the self-energy function $\Sigma(E)$ is
expressed in terms of the form factor
\andy{selfen}
\beq
\label{eq:selfen}
\Sigma(E)=\int d\omega\; \frac{\left|\bra{a}H_{\rm
int}\ket{\omega}\right|^2}{E-\omega} =\int
d\omega\;\frac{g^2(\omega)}{E-\omega}  .
\eeq
A straightforward analysis in terms of the resolvent of the
Hamiltonian yields
\andy{x2terms}
\beq
x(t)=\sqrt{{\cal Z}}e^{-\gamma_0t/2 -i\alpha(t)}+x_{\rm cut}(t) ,
\quad {\cal Z} = |1-\Sigma'(E_{\rm pole})|^{-2}
\label{x2terms}
\eeq
where the exponential term (first term) is due to the contribution of
a simple pole $E_{\rm pole}$ on the second Riemannian sheet in the
complex energy plane, while the second term is the result of a
contour integration
\cite{temprevi}. The lifetime $\gamma_0^{-1}$ is given by the Fermi
``golden" rule, computed according to the Weisskopf-Wigner
approximation. The quantity ${\cal Z}$ is the square of the residue
of pole of the propagator (wave function renormalization) and
$\alpha$ a (real) linear function of time. The cut contribution is of
order (coupling constant)$^2$ and modifies the exponential law both
at short and long times, yielding the characteristic quadratic and
power-law behaviors. The survival probability reads then
\andy{psurv}
\beq
\label{eq:psurv}
P(t)= |x(t)|^2 = {\cal Z} e^{-\gamma_0 t} + \mbox{other terms}.
\eeq
The above results are of general validity.

\begin{figure}
\centerline{\epsfig{file=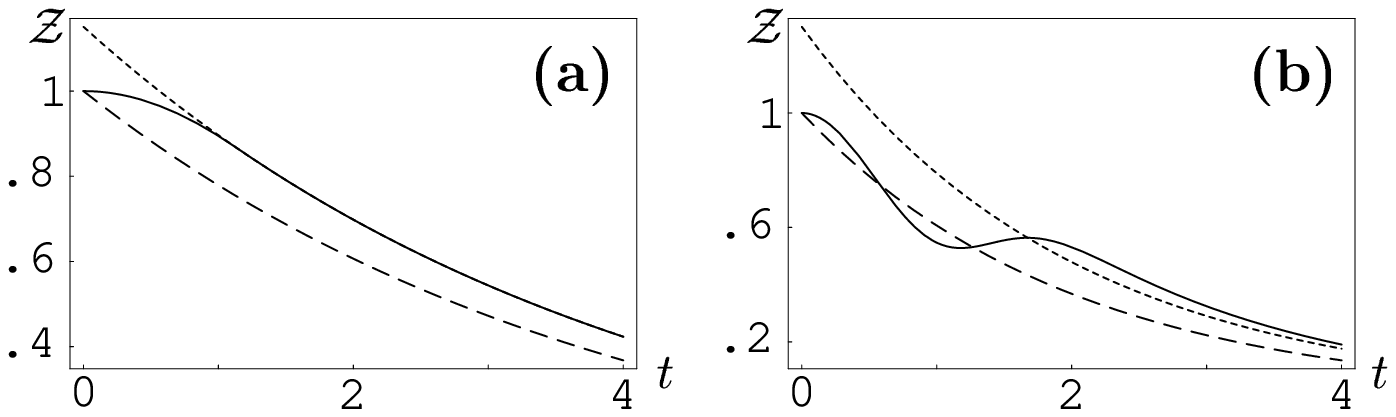,width=\figwidth}}
\caption{Study of the case ${\cal Z} > 1$. The full line is the
survival probability, the dashed line the renormalized exponential
$e^{-\gamma_0 t}$ and the dotted line the asymptotic exponential
${\cal Z} e^{-\gamma_0 t}$. (a) If $P(t)$ and $e^{-\gamma_0t}$ do not
intersect, a finite solution $\tau^*$ does not exist. (b) If $P(t)$
and $e^{-\gamma_0t}$ intersect, a finite solution $\tau^*$ exists.
(In this case there are always at least two intersections.) }
\label{fig:Zren}
\end{figure}

The following theorem holds: in general, a sufficient condition for
the existence of a solution $\tau^*$ of Eq.\ (\ref{eq:tstardef}) is
${\cal Z}<1$. The best proof of this proposition is obtained by
graphical inspection: The case ${\cal Z}<1$ is shown in Fig.\
\ref{fig:gtau}(a): $P(t)$ and $e^{-\gamma_0t}$ {\em must}
intersect, since according to (\ref{eq:psurv}), $P(t) \sim {\cal
Z}e^{-\gamma_0t}$ for large $t$ \cite{foot}, and a finite solution
$\tau^*$ can always be found. The other case, ${\cal Z} > 1$, is
shown in Fig.~\ref{fig:Zren}: a solution may or may not exist,
depending on the model. Interestingly, the above theorem shows that
renormalization plays an important role in the Zeno problem, when one
deals with unstable systems.

\begin{figure}
\centerline{\epsfig{file=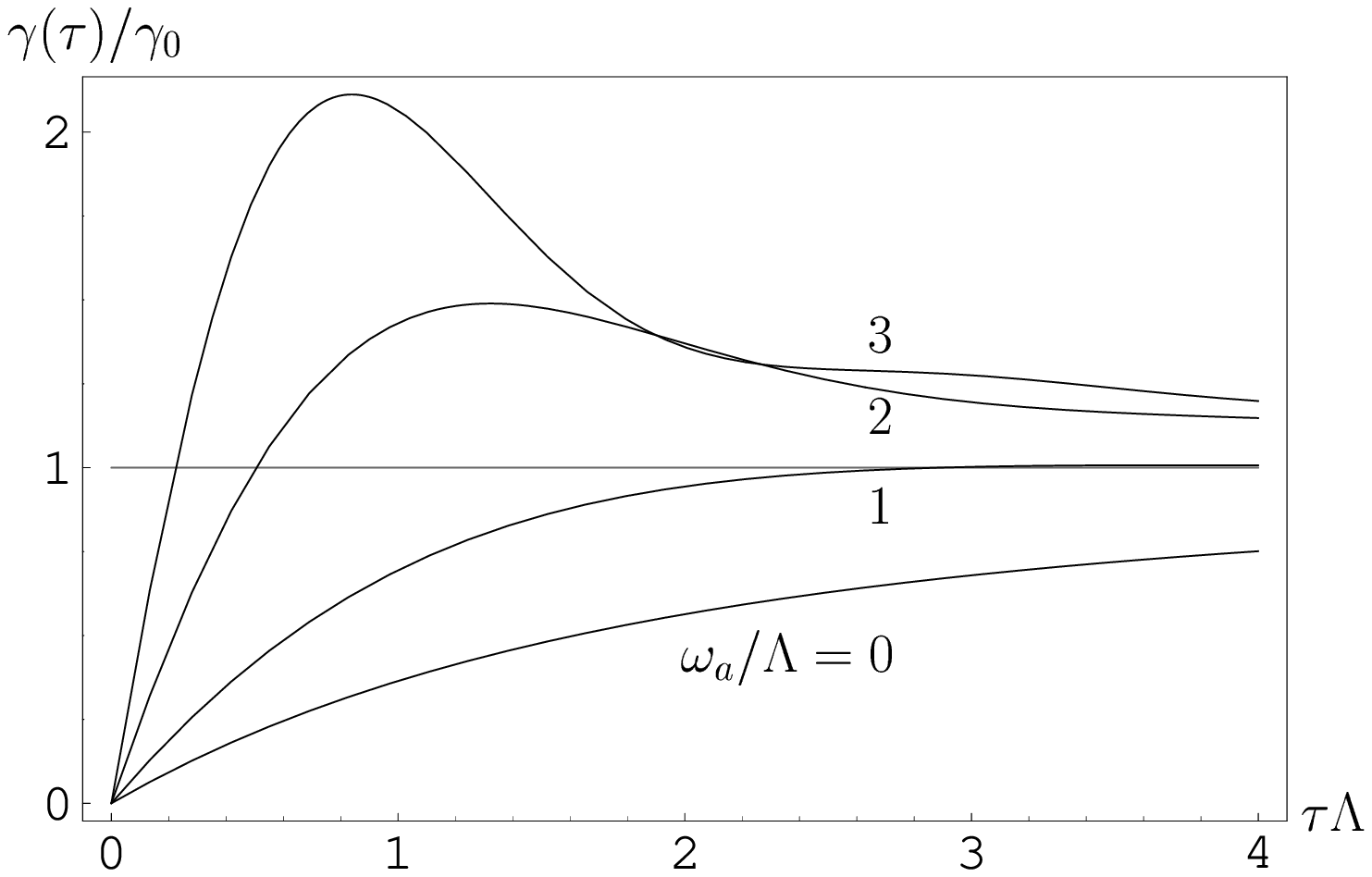,width=8.5cm}}
\caption{Effective decay
 rate $\gamma(\tau)$ for the model
(\ref{eq:2formfact}), for $\lambda=0.1$ and different values of
the ratio $\omega_a/\Lambda$ (indicated). The horizontal line
shows the ``natural" decay rate $\gamma_0$: its intersection with
$\gamma(\tau)$ yields the solution $\tau^*$ of Eq.\
(\ref{eq:tstardef}). The asymptotic value of all curves is
$\gamma_0$, as expected. A Zeno (inverse Zeno) effect is obtained
for $\tau<\tau^*$ ($\tau>\tau^*$). Notice the presence of a linear
region for small values of $\tau$ and observe that $\tau^*$ does
not belong to such linear region as the ratio $\omega_a/\Lambda$
decreases. Above a certain threshold, given by Eq.\
(\ref{eq:solZ}) in the weak coupling limit of the model (and in
general by the condition ${\cal Z}=1$), Eq.\ (\ref{eq:tstardef})
has no finite solutions: only a Zeno effect is realizable in such
a case.}
\label{fig:gammat}
\end{figure}

In order to check our general conclusions and investigate the primary
role played by the specific features of the interaction, let us first
focus on a Lorentzian form factor
\andy{2formfact}
\beq
\label{eq:2formfact}
g(\omega)=\frac{\lambda}{\sqrt{\pi}}\sqrt{\frac{\Lambda}{\omega^{2}+\Lambda^{2}}}.
\eeq
This describes, for instance, an atom-field coupling in a cavity with
high finesse mirrors \cite{cavity} and has the advantage of being
solvable. The role of form factors in the context of the QZE was
studied in earlier papers \cite{Lane,KK96,FPOlomouc}. In particular,
Kofman and Kurizki also considered the Lorentzian case. (We stress
that in this case the Hamiltonian is not lower bounded and we expect
no deviations from exponential behavior at very large times, since
Khalfin's argument
\cite{Khalfin57} is circumvented.) One easily obtains
$\Sigma(E)=\lambda^2/(E+i\Lambda)$, whence the propagator
$x(E)=i(E+i\Lambda)/[(E-\omega_a)(E+i\Lambda)-\lambda^2]$ has two
poles in the lower half energy plane and yields
\andy{xlor}
\beq
\label{eq:xlor}
x(t)=
\frac{\omega_a+\Delta+i(\Lambda-\gamma_0/2)}{\omega_a+2\Delta+i(\Lambda-\gamma_0)}
e^{-i(\omega_a+\Delta)t} e^{-\gamma_0 t/2}
+\frac{\Delta-i\gamma_0/2}{\omega_a+2\Delta+i(\Lambda-\gamma_0)}
e^{i\Delta t} e^{-(\Lambda-\gamma_0/2) t},
\eeq
where $\Delta=-\omega_a/2+(\omega_a/2)
\sqrt{(\sqrt{\Omega^4+4\omega_a^2\Lambda^2}+\Omega^2)/2\omega_a^2}$
and $\gamma_0=\Lambda-
 \sqrt{(\sqrt{\Omega^4+4\omega_a^2\Lambda^2}-\Omega^2)/2}$,
with $\Omega^2=\omega_a^2+4\lambda^2-\Lambda^2$. In this case the
wave function renormalization reads
\andy{zz}
\beq
\label{eq:zz}
{\cal Z} =
\frac{(\omega_a+\Delta)^2+(\Lambda-\gamma_0/2)^2}{(\omega_a+2\Delta)^2+(\Lambda-\gamma_0)^2}.
\eeq
By plugging (\ref{eq:xlor}) into (\ref{eq:gammadef}) one obtains the
effective decay rate, whose behavior is displayed in
Fig.~\ref{fig:gammat} for different values of the ratio
$\omega_a/\Lambda$.
These curves show that for large values of $\omega_a$ (in units
$\Lambda$) there is indeed a transition from a Zeno to an inverse
Zeno (``Heraclitus") behavior: such a transition occurs at
$\tau=\tau^*$, solution of Eq.\ (\ref{eq:tstardef}). However, for
small values of $\omega_a$, such a solution ceases to exist. The
determination of the critical value of $\omega_a$ for which the
Zeno--inverse Zeno transition ceases to take place discloses an
interesting aspect of this issue. The problem can be discussed in
general, but for the sake of simplicity we consider the weak coupling
limit (small $\lambda$): in this case the other terms in
(\ref{eq:psurv}), arising from the second addendum in
(\ref{eq:xlor}), are of order $\lambda^2$ and quickly vanish for
large $t$ ($\gamma_0$ is of order $\lambda^2$). Moreover, by
(\ref{eq:zz}) the inequality ${\cal Z}<1$ yields
\andy{solZ}
\beq
\label{eq:solZ}
\omega_a^2 > \Lambda^2 + \mbox{O}(\lambda^2).
\eeq
The meaning of this relation is the following: a sufficient condition
to obtain a Zeno--inverse Zeno transition is that the energy of the
decaying state be placed asymetrically with respect to the peak of
the form factor (bandwidth). If, on the other hand, $\omega_a \simeq
0$ (center of the bandwidth), no transition time $\tau^*$ exists (see
Fig.\ \ref{fig:gammat}) and only a QZE is possible: this is the case
analyzed in Fig.\ \ref{fig:Zren}(a). A relation similar to
(\ref{eq:solZ}) was also discussed in \cite{KK96}.

There is more: Equation (\ref{eq:xlor}) yields a time scale. Indeed,
from the definitions of the quantities in (\ref{eq:xlor}) one gets
$\gamma_0/2 < \Lambda - \gamma_0/2$, so that the second exponential
in (\ref{eq:xlor}) vanishes more quickly than the first one
\cite{stronc}. If the coupling is weak, since
$\gamma_0=\mbox{O}(\lambda^2)$, the second term is very rapidly
damped so that, after a short initial quadratic region of duration
$\Lambda^{-1}$, the decay becomes purely exponential with decay rate
$\gamma_0$. This is an important point, often misunderstood in the
literature: the quadratic behavior $P(t)
\simeq \exp (-t^2/\tau_{\rm Z}^2)$ is valid {\em not} for times
$t\lesssim \tau_{\rm Z}=\lambda^{-1}$, but rather for much shorter
times $t\lesssim \Lambda^{-1}$. For $\tau \lesssim 1/\Lambda$ (which
is, by definition, the meaning of ``short" times in a quantum Zeno
context), we can use the linear approximation
\beq
\label{eq:taushort}
\gamma(\tau) \simeq \frac{\tau}{\tau_{\rm
Z}^2}\quad\mbox{for}\quad\tau \lesssim 1/\Lambda,
\eeq
where $\tau_{\rm Z}^{-2} \equiv \bra{a} H_{\rm int}^2 \ket{a} =
\int d\omega g^2(\omega)$. When the linear approximation
(\ref{eq:taushort}) applies up to the intersection (i.e.,
$|\omega_a|\gg \Lambda$) then
 \andy{jump}
\beq
\label{eq:jump}
 \tau^* \simeq \gamma_0 \tau_{\rm Z}^2.
\eeq
When the linear approximation does not hold, the r.h.s.\ of the
above expression yields a lower bound to the transition time
(\ref{eq:tstardef}). The quantity $\gamma_0 \tau_{\rm Z}^2$ is
also relevant in different contexts and has been called ``jump
time" by Schulman \cite{contvspulsed}.

The conclusions obtained for the simple model
(\ref{eq:2formfact}) are of general validity. Indeed, the form of
the ``rotating wave" interaction Hamiltonian (\ref{eq:contham})
is a very general one \cite{Peres}. In general, in Eq.\
(\ref{eq:contham}), for any $g(\omega)$, we assume that
$\omega_a>\omega_g$, where $\omega_g$ is the ground energy of the
continuous spectrum, and regard $\omega$ as a collective index
that can include some discrete variables (such as polarization in
the case of photons), but must include at least a continuous one.
The integral over $\omega$ is then a short notation for a sum
over discrete quantum numbers and an integral over continuous
ones. The matrix elements of the interaction Hamiltonian depend
of course on the physical model considered. However, for
physically relevant situations, the interaction smoothly vanishes
for small values of $\omega-\omega_g$ and quickly drops to zero
for $\omega > \Lambda$, a frequency cutoff related to the size of
the decaying system and the characteristics of the environment.
This is true both for cavities \cite{cavity}, as well as for
typical EM decay processes in vacuum, where the bandwidth $\Lambda
\simeq 10^{14}-10^{18}$s$^{-1}$ is given by an inverse
characteristic length (say, of the order of Bohr radius) and is
much larger than the (``natural") inverse lifetime $\gamma_0\simeq
10^{7}-10^{9}$s$^{-1}$ \cite{formfactH}.

For form factors that are roughly symmetric, all the conclusions
drawn for the Lorentzian model remain valid. The main role is played
by the ratio $\omega_a/\Lambda$. In general, the asymmetry condition
(\ref{eq:solZ}) is satisfied if the energy $\omega_a$ of the unstable
state is sufficiently close to the threshold. In fact, from the
definition of the Zeno time $\tau_{\rm Z}$ one has
\beq
\frac{1}{\tau_{\rm Z}^2}=\int
d\omega\; g^2(\omega)=g^2(\bar\omega) \Lambda,
\eeq
where $\bar\omega$ is defined by this relation and is of order
$\omega_{\rm max}$, the energy at which $g(\omega)$ takes the maximum
value. For $\omega_a$  sufficiently close to the threshold $\omega_g$
one has $g(\bar\omega)\gg g(\omega_a)$, the time scale
$\gamma_0\tau_{\rm Z}$ is well within the short-time regime, namely
\beq
\gamma_0 \tau_{\rm Z}^2=\frac{2\pi
g^2(\omega_a)}{g^2(\bar\omega)\Lambda}\ll \frac{1}{\Lambda},
\eeq
where the Fermi golden rule $\gamma_0 = 2\pi g^2(\omega_a)$ has been
used, and therefore the estimate (\ref{eq:jump}) is valid.

On the other hand, for a system such that $\omega_a-\omega_g
\simeq\Lambda$ (or, better, $\omega_a \simeq \;$center of the
bandwidth), $\tau^*$ does not necessarily exist and usually {\it
only} a Zeno effect can occur. In this context, it is useful and
interesting to observe that the Lorentzian form factor
(\ref{eq:2formfact}) in (\ref{eq:contham}) yields, in the limit
$g^2(\omega)=\Lambda^2\delta(\omega-\omega_a)$, the physics of a
two level system. This is also true in the general case, for a
roughly symmetric form factor, when the bandwidth $\Lambda\to 0$.
In such a case, the physical conditions leading to QZE are
readily realizable \cite{Cook} (and no transition to IZE is
possible).

Some final comments are in order. The present analysis has been
performed in terms of instantaneous measurements, according to
the Copenhagen prescription. Our starting point was indeed Eq.\
(\ref{eq:effgamma}). We cannot help feeling that such a
formulation of the QZE is unsatisfactory, even in the simplest
case of two level systems \cite{BeigeHegerl}. A more exhaustive
formulation, that takes into account the state of the detection
system and the physical duration of the measurement process will
be presented elsewhere. This approach, performed in terms of
``continuous" measurements
\cite{Kraus,MPS,contvspulsed,FPOlomouc,perina} circumvents the
(very subtle) conceptual problem of state preparation, which
affects most field theoretical formulations of the QZE. It is also
worth emphasizing that the first experimental evidence of
non-exponential decay at short times is very recent \cite{Raizen}
and no hindered evolution due to repeated measurements (QZE) has
ever been observed for {\em bona fide} unstable systems. The
approach we propose might lead to new ideas for an experimental
verification of these effects.

\section*{Acknowledgements}
We thank L.S.\ Schulman for interesting comments.



\begin{thebibliography}{99}
\bibitem{foot} \andy{foot}
Namely, for times that are not too short (so that the cut
contribution has become negligible) and not too long (so that no
power low regime has taken over).
\end{thebibliography}
 \end{document}